\def\famname{
 \textfont0=\textrm \scriptfont0=\scriptrm
 \scriptscriptfont0=\sscriptrm
 \textfont1=\textmi \scriptfont1=\scriptmi
 \scriptscriptfont1=\sscriptmi
 \textfont2=\textsy \scriptfont2=\scriptsy \scriptscriptfont2=\sscriptsy
 \textfont3=\textex \scriptfont3=\textex \scriptscriptfont3=\textex
 \textfont4=\textbf \scriptfont4=\scriptbf \scriptscriptfont4=\sscriptbf
 \skewchar\textmi='177 \skewchar\scriptmi='177
 \skewchar\sscriptmi='177
 \skewchar\textsy='60 \skewchar\scriptsy='60
 \skewchar\sscriptsy='60
 \def\rm{\fam0 \textrm} \def\bf{\fam4 \textbf}}
\def\sca#1{scaled\magstep#1} \def\scah{scaled\magstephalf} 
\def\twelvepoint{
 \font\textrm=cmr12 \font\scriptrm=cmr8 \font\sscriptrm=cmr6
 \font\textmi=cmmi12 \font\scriptmi=cmmi8 \font\sscriptmi=cmmi6 
 \font\textsy=cmsy10 \sca1 \font\scriptsy=cmsy8
 \font\sscriptsy=cmsy6
 \font\textex=cmex10 \sca1
 \font\textbf=cmbx12 \font\scriptbf=cmbx8 \font\sscriptbf=cmbx6
 \font\it=cmti12
 \font\sectfont=cmbx12 \sca1
 \font\sectmath=cmmib10 \sca2
 \font\sectsymb=cmbsy10 \sca2
 \font\refrm=cmr10 \scah \font\refit=cmti10 \scah
 \font\refbf=cmbx10 \scah
 \def\twelverm{\textrm} \def\twelveit{\it} \def\twelvebf{\textbf}
 \famname \textrm 
 \advance\voffset by.2in \advance\hoffset by.3in
 \normalbaselineskip=17.5pt plus 1pt \baselineskip=\normalbaselineskip
 \parindent=21pt
 \setbox\strutbox=\hbox{\vrule height10.5pt depth4pt width0pt}}

\overfullrule=0pt			
\vsize=9in \hsize=6in
\lineskip=0pt				
\abovedisplayskip=1.2em plus.3em minus.9em 
\belowdisplayskip=1.2em plus.3em minus.9em	
\abovedisplayshortskip=0em plus.3em	
\belowdisplayshortskip=.7em plus.3em minus.4em	
\def\makefootline{\baselineskip=30pt \line{\the\footline}}
\footline={\ifnum\count0=1 \hfil \else\hss\twelverm\folio\hss \fi}
\pageno=1

\def\center{\leftskip=0pt plus 1fill \rightskip=\leftskip \parindent=0pt
 \def\textindent##1{\par\hangindent21pt\footrm\noindent\hskip21pt
 \llap{##1\enspace}\ignorespaces}\par}
\def\unnarrower{\leftskip=0pt \rightskip=\leftskip}

\def\thetitle#1#2#3#4#5{
 \font\titlefont=cmbx12 \sca2 \font\footrm=cmr10 \font\footit=cmti10
  \twelverm
	{\hbox to\hsize{#4 \hfill YITP-SB-#3}}\par
	\vskip.8in minus.1in {\center\baselineskip=1.44\normalbaselineskip
 {\titlefont #1}\par}{\center\baselineskip=\normalbaselineskip
 \vskip.5in minus.2in #2
	\vskip1.4in minus1.2in {\twelvebf ABSTRACT}\par}
 \vskip.1in\par
 \narrower\par#5\par\unnarrower\vskip3.5in minus2.3in\eject}
\def\paper\par#1\par#2\par#3\par#4\par#5\par{\twelvepoint
	\thetitle{#1}{#2}{#3}{#4}{#5}} 
\def\author#1#2{#1 \vskip.1in {\twelveit #2}\vskip.1in}
\def\YITP{C. N. Yang Institute for Theoretical Physics\\
	State University of New York, Stony Brook, NY 11794-3840}

\def\itemize#1 {\item{[#1]}} 


\newcount\marknumber	\marknumber=1
\newcount\countdp \newcount\countwd \newcount\countht 

%
%
\ifx\pdfoutput\undefined
\def\rgboo#1{}
\input epsf

\def\figscale#1#2{\epsfxsize=#2\epsfbox{#1.eps}}
\def\postscript#1{\special{" #1}}		
\postscript{
	/bd {bind def} bind def
	/fsd {findfont exch scalefont def} bd
	/sms {setfont moveto show} bd
	/ms {moveto show} bd
	/pdfmark where		
	{pop} {userdict /pdfmark /cleartomark load put} ifelse
	[ /PageMode /UseOutlines		
	/DOCVIEW pdfmark}
\def\bookmark#1#2{\postscript{		
	[ /Dest /MyDest\the\marknumber /View [ /XYZ null null null ] /DEST pdfmark
	[ /Title (#2) /Count #1 /Dest /MyDest\the\marknumber /OUT pdfmark}%
	\advance\marknumber by1}
\def\pdfklink#1#2{%
	\hskip-.25em\setbox0=\hbox{#1}%
		\countdp=\dp0 \countwd=\wd0 \countht=\ht0%
		\divide\countdp by65536 \divide\countwd by65536%
			\divide\countht by65536%
		\advance\countdp by1 \advance\countwd by1%
			\advance\countht by1%
		\def\linkdp{\the\countdp} \def\linkwd{\the\countwd}%
			\def\linkht{\the\countht}%
	\postscript{
		[ /Rect [ -1.5 -\linkdp.0 0\linkwd.0 0\linkht.5 ] 
		/Border [ 0 0 0 ]
		/Action << /Subtype /URI /URI (#2) >>
		/Subtype /Link
		/ANN pdfmark}{\rgb{1 0 0}{#1}}}
%
%
\else
\def\rgboo#1{\pdfliteral{#1 rg #1 RG}}

\def\figscale#1#2{\pdfximage width#2 {#1.pdf}\pdfrefximage\pdflastximage}
\pdfcatalog{/PageMode /UseOutlines}		
\def\bookmark#1#2{
	\pdfdest num \marknumber xyz
	\pdfoutline goto num \marknumber count #1 {#2}
	\advance\marknumber by1}
\def\pdfklink#1#2{%
	\noindent\pdfstartlink user
		{/Subtype /Link
		/Border [ 0 0 0 ]
		/A << /S /URI /URI (#2) >>}{\rgb{1 0 0}{#1}}%
	\pdfendlink}
\fi

\def\rgbo#1#2{\rgboo{#1}#2\rgboo{0 0 0}}
\def\rgb#1#2{\mark{#1}\rgbo{#1}{#2}\mark{0 0 0}}
\def\pdflink#1{\pdfklink{#1}{#1}}
\def\xxxlink#1{\pdfklink{#1}{http://arXiv.org/abs/#1}}

\catcode`@=11

\def\wlog#1{}	


\def\makeheadline{\vbox to\z@{\vskip-36.5\p@
	\line{\vbox to8.5\p@{}\the\headline%
	\ifnum\pageno=\z@\rgboo{0 0 0}\else\rgboo{\topmark}\fi%
	}\vss}\nointerlineskip}
\headline={
	\ifnum\pageno=\z@
		\hfil
	\else
		\ifnum\pageno<\z@
			\ifodd\pageno
				\tenrm\romannumeral-\pageno\hfil\lefthead\hfil
			\else
				\tenrm\hfil\righthead\hfil\romannumeral-\pageno
			\fi
		\else
			\ifodd\pageno
				\tenrm\hfil\righthead\hfil\number\pageno
			\else
				\tenrm\number\pageno\hfil\lefthead\hfil
			\fi
		\fi
	\fi}

\catcode`@=12

\def\righthead{\hfil} \def\lefthead{\hfil}
\nopagenumbers


\def\sectskip{\par\ifdim\lastskip<\medskipamount
	\bigskip\medskip\goodbreak\else\nobreak\fi}
\def\secty#1{
	\xdef\righthead{\rgbo{1 0 1}{#1}}\vbox{\bigger\offinterlineskip
	\line{\cdotfill¼\hphantom{\righthead}\cdotfill}
	\line{\cdotfill¼\righthead\cdotfill}}\par\nobreak\medskip}
\def\sect#1 #2\par{\sectskip\bookmark{#1}{#2}\secty{#2}}
\def\subsectskip{\par\ifdim\lastskip<\medskipamount
	\bigskip\medskip\goodbreak\else\nobreak\fi}
\def\subsecty#1{\noindent{\sectfont{\rgbo{.5 0 1}{#1}}}\par\nobreak\medskip}
\def\subsect#1\par{\subsectskip\bookmark0{#1}\subsecty{#1}}
\def\refs{\bigskip\noindent{\bf \rgbo{0 .5 1}{REFERENCES}}\par\nobreak\medskip
	\frenchspacing \parskip=0pt \let\it=\refit \let\rm=\refrm \rm
	\baselineskip=1.23em plus 1pt}

\def\\{\hfil\break}
\def\half{{\textstyle{1\over{\raise.1ex\hbox{$\scriptstyle{2}$}}}}}
\def\frac#1#2{{\textstyle{#1\over#2}}}
\def\on#1#2{{\buildrel{\mkern2.5mu#1\mkern-2.5mu}\over{#2}}}
\def\ron#1#2{{\buildrel{#1}\over{#2}}}
\def\bop#1{\setbox0=\hbox{$#1M$}\mkern1.5mu
	\vbox{\hrule height0pt depth.04\ht0
	\hbox{\vrule width.04\ht0 height.9\ht0 \kern.9\ht0
	\vrule width.04\ht0}\hrule height.04\ht0}\mkern1.5mu}
\def\bo{{\mathpalette\bop{}}}                        

\font\bigc=cmssdc10 scaled 2550

\pageno=0

\paper

\bigc{\rgb{1 0 .5}{Bound-state gravity from higher derivatives}}

\author{Kiyoung Lee\footnote{${}^1$}{
	\pdflink{mailto:klee@grad.physics.sunysb.edu}}
	and Warren Siegel\footnote{${}^2$}{
	\pdflink{mailto:siegel@insti.physics.sunysb.edu}}}\YITP

03-12

March 19, 2003

In certain Lorentz-covariant higher-derivative field theories of spins $\le$1, would-be ultraviolet divergences generate color-singlet poles as infrared divergences.  Absence of higher-order poles implies ten-dimensional supersymmetric Yang-Mills with bound-state supergravity, in close analogy with open string theory.

\subsect Introduction

Many properties discovered in string theory have been reproduced in ordinary field theory: supersymmetry, the topological (1/N) expansion, the first-quantized BRST approach to arbitrary gauge theories [1], the Gervais-Neveu gauge [2], and certain simplifications in one-loop amplitudes [3].  In that respect string theory has proven a useful toy model.  The main properties of string theory that have not been reproduced so far in particle field theory are:  

(1) Dolen-Horn-Schmid duality (duality between different momentum channels).  This is the reason ``dual models" were first developed.  It is also the only experimentally verified property of strings, as applied to hadrons (along with its direct consequences).  

(2) Theories of fundamental higher-spin particles with true predictive power.  Known string theories are perturbatively renormalizable, and are thus expected to predict more than possible from just unitarity, Poincar\'e invariance, locality, and low-energy symmetries.  However, it is not clear how much of this predictability survives compactification and other nonperturbative effects (such as the appearance of even more dimensions, from M- or F-theory).  

(3) Higher dimensions of spacetime.  This may apply only to the string models that have been explicitly constructed so far, and not to hadronic strings (although attempts have been made through the AdS/CFT correspondence [4]).  Extended supergravity theories hint at higher dimensions, but do not require them.  This property requires some predictive theory of gravity to hide the extra dimensions.

(4) Gravitons appear as bound states in theories without fundamental massless spin-two states (open strings).  This is actually only a kinematic effect (analogous to bosonization in two-dimensional theories) [5], but is the only known explicit example of the generation of the graviton as a bound state.  (Long after this effect was discovered, a ``theorem" was discovered that indicated its impossibility [6].  A proposed explanation was an anomaly in the conservation of the energy-momentum tensor [7], since the essence of the theorem is the discrepancy between the ordinary conservation law of nongravitational theories and the covariant conservation law in curved space.)  Confinement in QCD would also allow derivation of higher-spin states as bound states, and perhaps without loss of predictive power (if ambiguities presented by renormalons can be overcome, perhaps by embedding in finite theories), but explicit calculational methods (lattice QCD and various nonlinear sigma-model methods) so far have proven inadequate to examine this effect.  Random lattice models of strings [8] may provide an example of confinement, but so far explicit calculations have been limited to lower dimensions.  Of course, if an ordinary field theory can be shown to reproduce this property of higher spin as bound states, the second property above should become irrelevant.

In this paper we provide an example of a field theory that reproduces the latter two properties.  This theory is a Lorentz-covariant higher-derivative modification of ten-dimensional supersymmetric Yang-Mills theory, the massless sector of the open superstring.

\subsect Open string theory

Any one-loop string theory diagram composed completely of open strings contains closed strings, as a consequence of duality (the ``stretchiness" of the world sheet):  Any such (orientable) diagram can be made planar by allowing external lines to extend from either the outside or inside of the loop.  Then the closed string propagates between the inner and outer boundaries of the loop, which can be stretched apart to give a cylinder.  (In the case where all states are on one boundary, the other boundary represents the closed-string connecting to the vacuum.)  As internal symmetry (``color") is associated with the boundaries, the closed string is a color singlet.  Self interactions of closed strings are similarly found in higher loops of open strings, as follows from the usual relation $g_{closed}=\hbar g_{open}^2$ between the fundamental coupling of the open string and the effective coupling of the closed-string composite state.

$$ \figscale{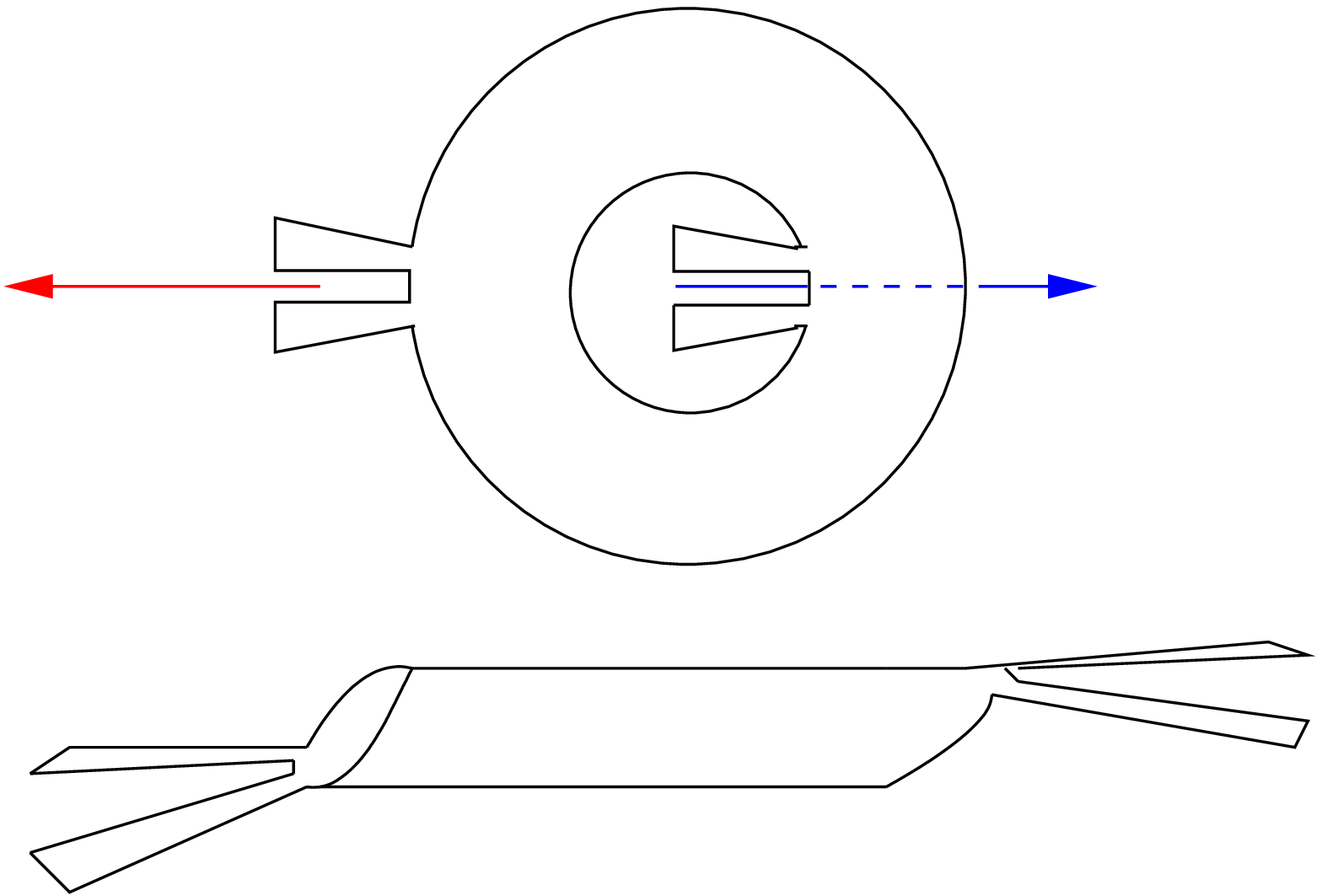}{3in} $$

The same stretchiness allows such one-loop diagrams to be drawn as a one-loop open-string propagator correction with a half-twist in each side of the loop, and all the external states of either boundary clustered at the end of either propagator extending from the loop.  Thus the generation of the closed string can be associated with the propagator correction itself.  But any (amputated) one-loop propagator correction is equivalent to the expectation of the product of two quadratic operators in the free theory, where the vertices of the correction serve to define the operators.  Thus any bound state appearing in a one-loop propagator correction (or any propagator correction with only two vertices) can be recognized as a kinematic effect.  (A familiar example is the massless scalar generated from currents of massless spinors in two dimensions, which is then eaten by the vector in the Schwinger model.)  Thus string theory is not an ideal example of generation of the graviton as a bound state.  Yet it might share enough characteristics with true bound-state models to provide some insight.  

Furthermore, random lattice models of string theory explicitly represent strings as generated from resummation of Feynman diagrams of particle field theory (e.g., scalar theories for the bosonic string), in a manner expected for the hadronic string from QCD.  However, these models have Gaussian propagators rather than poles (although it might be possible to generalize this construction to more realistic models [9]).

There are several features of string theory with which we could try to account for this phenomenon:  (1) Unbounded mass.  Strings have particles of the same type but arbitrarily large mass, at integer spacing in (mass)${}^2$.  (2) Unbounded spin.  Strings have particles of the same type but arbitrarily large spin, at integer spacing.  (The maximum spin depends on the mass.)  (3) Unbounded derivatives.  Any given state couples with arbitrarily high derivatives.  (The minimum number depends on the spin.)

We now examine this one-loop feature of string theory in more detail in an attempt to isolate the mechanism responsible.  In superstring theory the one-loop integral can be written in the form (after all loop momenta have been integrated out)
$$ A_{string} \approx \int_0^\infty d(1/\tau)\ e^{\pi^2 \alpha'^2 s/\tau} \sim
 {1\over \alpha'^2 s} $$
where $\tau$ is the sum of the Schwinger parameters $\tau_i$ of the propagators running around the loop ($1/k_i^2=\int_0^\infty d\tau_i\ e^{-\tau_ik_i^2}$), related to the usual string parameter $q$ as $q=e^{-2\pi^2 \alpha'/\tau}$, ``$\approx$" means neglecting a factor in the integrand analytic in $q$ (we look at the contribution near $q=0$, $\tau=0$, normally associated with UV divergences, to isolate just the massless pole), and $s=-(\sum p)^2$ (summed over the momenta for the lines of either boundary) is the invariant for the closed-string state.  (We have also neglected an overall factor polynomial in external momenta, required for gauge invariance of external fields, etc.  For the bosonic string $s$ is shifted by a constant, giving the tachyon pole.)  Note that this result is nonperturbative in $\alpha'$.  If we expand the integrand in $\alpha'$ with hopes of resummation, we see terms of increasing UV divergence:  By dimensional analysis, increasing powers of loop momentum $k^2$ give decreasing powers of $\tau$.
Thus, higher-derivative couplings are the feature of string theory responsible for generation of the bound state at one loop.  (Higher spin might reproduce this effect, but only by virtue of higher derivatives.)

In dimensions other than $D=10$ there is an extra factor of a power of $e^\tau$ (e.g., $e^{\tau(26-D)/24\alpha'}$ for the bosonic string).  In field theory a massive particle in a loop will generate an extra factor $e^{-m^2 \tau}$.  We are thus led to consider only massless fundamental particles in order to avoid the usual problem in noncritical string theory of generating closed strings whose ``free" propagators contain cuts in addition to poles.

\subsect Scalars

The higher-derivative coupling most resembling those in string field theory is a ``proper-time cutoff", including a factor $e^\bo $ on each field at a vertex.  However, such a coupling regularizes loop integrals, and would thus eliminate the UV divergence that should be the source of the bound state.  (We do not modify the propagator, to avoid problems with unitarity.)  However, a similar modification produces the desired result:  For example, for a scalar matrix field, we consider a Lagrangian of the form
$$ L \sim 
-\phi\bo \phi +\phi(e^{-i\pi\alpha'\bo /2}\phi)(e^{i\pi\alpha'\bo /2}\phi) $$
with $S=(tr\ L)/g^2$.  (The normalization of $\alpha'$ has been chosen to make the result agree with string theory.)  We will consider only U(N) matrices, with the usual geometric correspondence to open string diagrams via the 't Hooft double-line notation.  

$$ \figscale{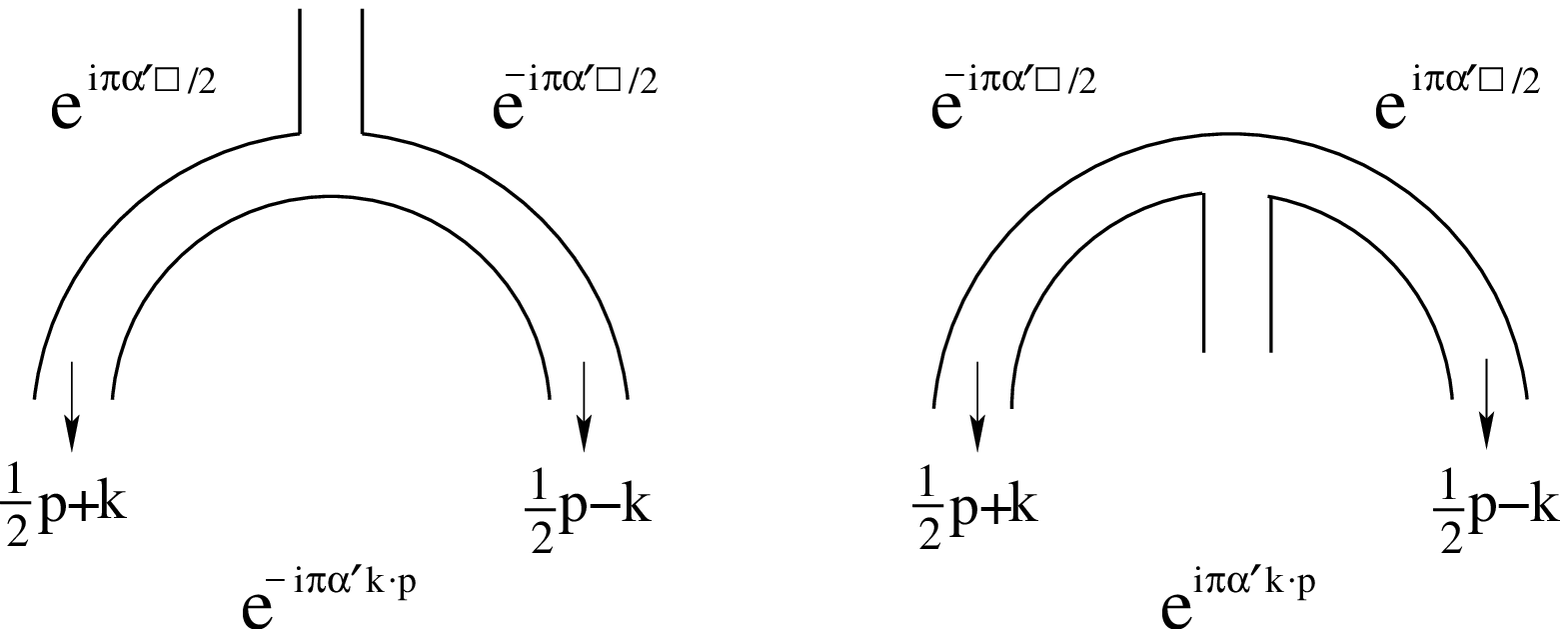}{4in} $$

If both the fields with higher-derivative operators contribute to the internal lines of a one-loop graph, then a vertex factor of the form $e^{\pm i\pi\alpha'k\cdot p}$ will result, where $k$ is the loop momentum and $p$ the external momentum, up to a factor dependent only on $p^2$.  (If one higher-derivative factor acts on an external line, there will be an exponential factor with a $k^2$, which we will interpret as regularizing the graph, making it uninteresting, ignoring any ambiguities associated with the sign of the exponent.  A similar vertex factor of the form $e^{-\alpha'^2 k^2 p^2}$ has also been considered [10], but does not seem to reproduce the desired results.)  The sign of the exponential depends on which higher derivative acts on which internal line, and is directly correlated to the group-theory factor coming from the matrices.  Making the usual interpretation of the group-theory double lines with the ends of the string, then in string language we get one sign for external lines attached to one boundary, and the opposite for the other.  (The same result can be obtained by noting cancelation of exponentials from opposite ends of untwisted propagators, so only twisted ones contribute.)

The net result will then be an integral of the form, at high ``energy" $k^2$,
$$ A_{particle} \approx\int{d^D k\over (k^2)^n} e^{\pm i2\pi\alpha'k\cdot P} 
\sim (P^2)^{n-D/2} $$
where $\pm P$ is the total momentum of states on either boundary, and $n$ is the number of external lines.  To see a bound state in the S-matrix, we need to look at a four-point graph:  We therefore need $D=10$ to get a simple pole in that graph.  As for the string, a naive quadratic divergence is converted into a pole, here by an integral resembling a Fourier transform.  For comparison to the string theory result, we introduce Schwinger parameters:  After integrating over loop momentum (and ignoring the remaining Feynman parameter integrals, where $\tau_i=\tau\alpha_i$, as for the string above), we have (for $n=4$ in $D=10$) near $s=0$ (whose main contribution comes from $\tau\approx 0$, and thus large $k^2$) the same result as for strings,
$$ A_{particle} \approx \int_0^\infty d(1/\tau)\ e^{\pi^2 \alpha'^2 s/\tau} 
\sim {1\over \alpha'^2 s} $$

There is a relation of this coupling to higher-spin coupling:  Writing
$$ (e^{-i\pi\alpha'\bo /2}\phi)(e^{i\pi\alpha'\bo /2}\phi) =
\phi e^{i\pi\alpha'\ron\leftrightarrow\bo /2}\phi $$
and expanding in powers of $\bo$, we see each term is the (multiple) divergence of the higher-spin current
$$ \phi\on\leftrightarrow\partial{}^a ...\on\leftrightarrow\partial{}^b\phi $$
so such a coupling might be found in string theory if only the scalar Stueckelberg pieces of higher-spin fields are kept.  The single field used in particle theory might then be a linear combination of string fields of the same spin but different mass.  Note that these currents alternate in symmetry between the two $\phi$'s for increasing spin, in analogy to the group theory in string theory for symmetry groups SO(N) or USp(2N), where the representations alternate in symmetry between the two defining-representation indices for increasing mass, or spin of the leading trajectory.

There is a resemblance between this theory and noncommutative field theory:  If we define a (nonassociative) product and commutator, inspired by the above vertex, by
$$ A*B \equiv (e^{-i\pi\alpha'\bo /2}Ae^{i\pi\alpha'\bo /2})
(e^{i\pi\alpha'\bo /2}Be^{-i\pi\alpha'\bo /2}),\quad\quad
[A,B] \equiv A*B-B*A $$
then we find
$$ [x^a,x^b] \sim i\alpha'L^{ab} $$
in terms of the orbital angular momentum $L^{ab}=x^{[a}p^{b]}$, as in the original form of noncommutative field theory [11].  Furthermore, the conversion of a UV singularity into an IR one, and the explicit integrals, are similar to the Lorentz-noncovariant version of noncommutative field theory [12], where they also resemble closed-string states [13].  (However, other interpretations have been proposed [14].  The main focus in noncommutative field theory papers has been on renormalizable theories in $D=4$, whereas here we are mainly concerned with gravity, and reproducing the unitary and Poincar\'e covariant results of string theory, and so are led to $D=10$.  It is easy to see that similar considerations there would also lead uniquely to $D=10$, even though a generally covariant theory of gravity cannot be expected from a theory that is not even Lorentz covariant.)

There are a couple of problems with this scalar theory:  (1) The bound state has spin 0, which is not very interesting.  (2) Although the condition $D=10$ produces the desired pole in the four-point amplitude, lower-point diagrams have higher-order poles, which have no physical interpretation.  (One might expect to generate a pole from a higher-derivative quartic-interaction theory in a propagator-correction-like four-point graph.  However, a bound-state pole would not appear in such a graph, since in the corresponding field theory where that state is fundamental, with coupling $\bo h=T$, we find $\langle TT\rangle=\langle (\bo h)(\bo h)\rangle=\bo (1/\bo )\bo =\bo $, where $\bo $ is the kinetic operator, quadratic in derivatives.)  The solution to both these problems is Yang-Mills (+ lower spins), which will produce spin-2 (+ lower) bound states, as well as allow cancelation of the spurious poles in lower-point graphs (as in string theory) by judicious choice of matter.

\subsect Yang-Mills

The simplest way to evaluate one-loop graphs for external Yang-Mills is to use the background field gauge for internal Yang-Mills, and by parity-doubling internal spinors (squaring the kinetic operator).  The latter will have the effect of dropping Levi-Civita terms, but these require at least $D$ gamma-matrices, and thus at least $D/2$-point graphs, and for reasons explained above we focus on $D=10$ (but see remarks below) and no higher than four-point.  The generic kinetic operator for minimal coupling is then of the form $\bo +i F^{ab}S_{ab}$, where $\bo$ is background covariantized and $F^{ab}$ is the background field strength, while $S_{ab}$ is the spin operator.

In our case, rather than propose a higher-derivative generalization of the full action, we will propose one for this background-field action, which is only quadratic in the quantum fields (but arbitrary order in the Yang-Mills background).  For our generalization, we first rewrite this part of the minimally coupled Lagrangian as
$$ L \sim tr [ \psi ( \bo +2i F^{ab}S_{ab} ) \psi ] $$
where $\psi$ is spin 0, 1/2, or 1, and the factor of 2 comes from replacing the usual commutator for $F$ (when the quantum fields are written as matrices) with just a product.  This is equivalent here because of graded symmetry in the quantum fields together with the opposite graded symmetry for (the matrix indices of) the spin operators, but will not be true when we introduce the nonminimal higher derivatives.  (The covariant derivatives in $\bo$ keep the usual commutators.)

Then our naive higher-derivative generalization is
$$ L \sim 
tr [ (e^{i\pi\alpha'\bo /2}\psi) ( \bo +2i F^{ab}S_{ab} )(e^{-i\pi\alpha'\bo /2}\psi)] $$
Again, we will not derive this from a complete action, but we now explain how such higher derivative terms might be obtained from a gauge invariant action when $\psi$ is the vector.  There are several possibilities:\\  
(1) Gauge transformations themselves might be generalized to higher derivatives, as for noncommutative field theory.\\  
(2) Higher-derivative background-gauge fixing terms might automatically give such a form.\\  
(3) The $\bo$'s in the exponential might be replaced with quantum gauge covariant derivatives, turned into $\bo$ by the usual gauge fixing condition:  Writing
$$ (e^{-i\pi\alpha'\bo /2}A)(e^{i\pi\alpha'\bo /2}A) = $$
$$ [(e^{-i\pi\alpha'\bo /2}-1)A][(e^{i\pi\alpha'\bo /2}-1)A]
+A(e^{i\pi\alpha'\bo /2}A) +(e^{-i\pi\alpha'\bo /2}A)A -A^2 $$
the first term contains only $\bo$ terms, which can be replaced as 
$\bo A^a\to\nabla_b F^{ba}$ up to gauge fixing, the next two contribute only ``regularized" terms, which can be dropped, and the last gives the usual (after insertion of the minimal kinetic operator).

We now consider the conditions on the number of spin-0 and spin-1/2 states needed to cancel spurious poles arising at one loop.  Since these states have already been restricted to adjoint by group-theory considerations (e.g., defining representations will not cancel nonplanar singularities), cancelations in the two-point and three-point 1PI graphs reproduce the usual result of maximal supersymmetry (or at least its field content), for any $D$:  (1) Cancelation of the graphs without spin coupling $FS$ requires equal numbers of bosons and fermions (which then cancel for arbitrary numbers of external lines).  (2) Graphs with one spin coupling vanish trivially ($tr\ S=0$).  (3) Cancelation of graphs with two spin couplings (again for arbitrary external lines) then fixes the number of spin 1/2 fields:  Since $tr(SS)$ gives the SO(D) Casimir as usual, this condition fixes the number of spinors to be that of maximally supersymmetric Yang-Mills.  (4)  The graphs for three spin couplings can be divided into $tr(S[S,S])$ and $tr(S\{S,S\})$ terms.  The former reduce to $tr(SS)$, and so cancel by the previous condition.  The latter are of the same form as for 4D anomalies for SO(D) internal symmetry.  We would miss such contributions because of our parity doubling.  However, these Levi-Civita terms can appear only for $D=2$ or 6, and in our case only for chiral spinors.  Their cancelation then requires equal numbers of spinors of each chirality for those dimensions, as also required for maximally supersymmetric Yang-Mills in those cases.  (This would also justify parity doubling for general graphs, except for the case $D=10$, which unfortunately is the case of most interest.)  (5) The total result for the four-point graph is then given completely by spin couplings, and is thus given by the result of the $\phi^3$ calculation of the previous section times four $F$'s, with index contractions from $tr(S^4)$.  

$$ \vcenter{\hbox{\figscale{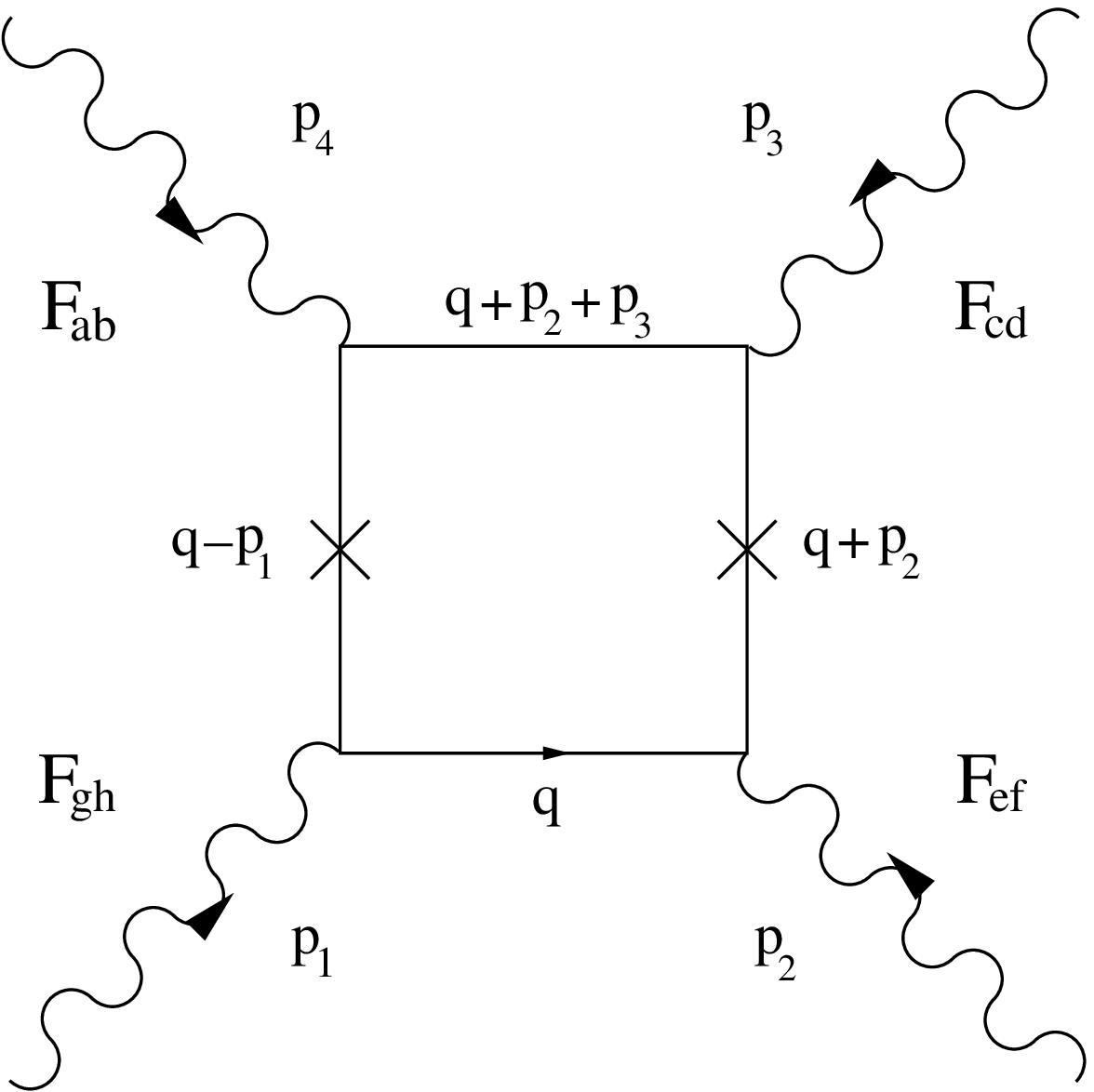}{2.5in}}} \quad\quad\to\quad\quad 
\vcenter{\hbox{\figscale{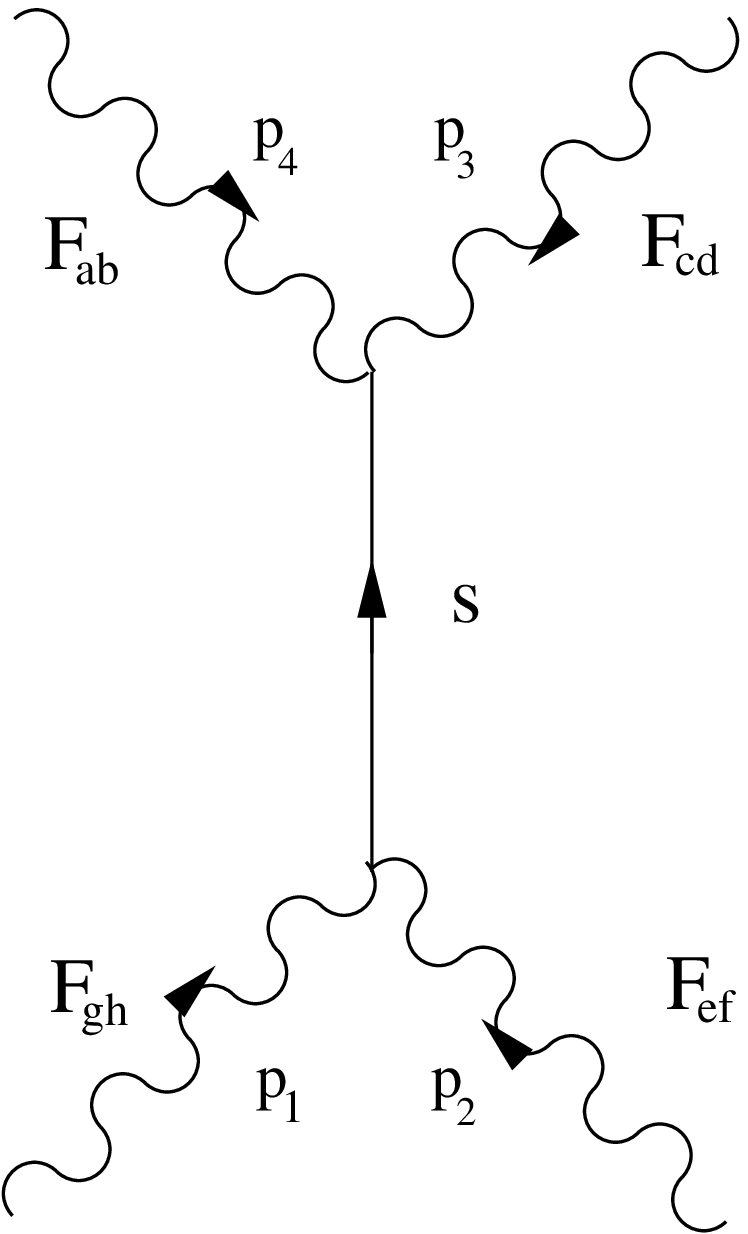}{1.5in}}} $$

Therefore $D=10$ is again required, yielding the same result as the zero-slope limit (but {\it not} $\alpha'=0$ identically) of string theory near the bound-state pole.  The required theory is thus a higher-derivative generalization of $10D$ supersymmetric Yang-Mills, which we assume preserves the supersymmetry.  We have not calculated the diagrams with external fermions, or even given an explicitly supersymmetric higher-derivative action with which to calculate them, but supersymmetry would require the bound states to be those of ($N=1$) supergravity.  

In this calculation (as in string theory) we can see the generation of the graviton, dilaton, and axion:  The result for the diagram with pairs of vertices on the ``inside" or ``outside" of the loop adjacent is of the form
$$ \Gamma \sim tr(F^{ab}F^{cd}) \frac1\bo tr(F^{ef}F^{gh})
	str(S_{ab}S_{cd}S_{ef}S_{gh}) $$
where the first two traces are for internal symmetry and the last for spin, which we have written as a ``supertrace" because we subtract the contribution of the fermion.  (There is also a factor of $\half$ for the fermion trace if we use it doubled.)  The vector spin operator is $S_{(1)ab}=|{}_{[a}\rangle\langle{}_{b]}|$, so its contribution will have an index contraction within each of the two internal-trace factors, and thus contribute only to the graviton and dilaton couplings.  Because of the cyclic symmetry of traces, the internal traces are symmetric in the two $F$'s, so the spin trace is symmetrized over the former pair of $S$'s and the latter.  Consequently the product of each pair of $S_{(1/2)ab}\sim \gamma_{[a}\gamma_{b]}$ will contain as matrices only the identity (times a pair of $\eta$'s) and $\gamma_{[a}\gamma_b\gamma_c\gamma_{d]}$, giving coupling to only the dilaton and the axion.  The axion is normally written as a two-form $B_{ab}$, but here it's convenient to think of it in terms of the dual four-form $B_{abcd}$ [15]:  At the present order in couplings, we can write for this form
$$ L \sim \frac1{12}(H_{abc})^2, \quad\quad H_{abc} = \half\partial_{[a}B_{bc]} +
	tr(\half A_{[a}\partial_b A_{c]} +\frac13 A_{[a}A_b A_{c]}) $$
$$ \to -\frac1{12}(H_{abc})^2 +\frac18 B^{abcd}tr(F_{ab}F_{cd}), \quad\quad 
	H^{abc} = \partial_d B^{abcd} $$
where the transformation is achieved as usual by switching field equations and Bianchi identities (e.g., by using a first-order form of the Lagrangian).

There is also a diagram with opposite vertices both on the inside or both outside, corresponding to switching the middle two $S$'s in the above trace.  This could give couplings other than those attributed to the graviton, dilaton, and axion, but the condition of cancelation between spins of all lower traces of $S$'s guarantees the symmetry of the total spin trace under all permutations of $S$'s, since anticommutators of two adjacent $S$'s replace them with structure constants.  This permutation symmetry alone is sufficient to fix the relative normalization of the graviton, dilaton, and axion contributions, which can then be written as
$$ \Gamma \sim tr(F^{ac}F^b{}_c)\frac1\bo tr(F_a{}^d F_{bd})
	-\frac18 tr(F^{ab}F_{ab})\frac1\bo tr(F^{cd}F_{cd}) $$
$$ -\frac14 tr(F^{ab}F^{cd})\frac1\bo tr(F_{ab}F_{cd}-2F_{ac}F_{bd}) $$
The first two terms are a linear combination of the graviton and dilaton terms, which can be explicitly separated as (for $D=10$)
$$ T^{ab}\frac1\bo T_{ab}
	-\frac14 tr(F^{ab}F_{ab})\frac1\bo tr(F^{cd}F_{cd}), \quad\quad
	T_{ab} = tr(F_a{}^c F_{bc} -\frac14 \eta_{ab}F^{cd}F_{cd}) $$
while for the axion terms we have used the antisymmetry of $F$ and the symmetry of the trace to reduce the number of terms in the manifestly antisymmetric form
$$ -\frac1{4!32}tr(F^{[ab}F^{cd]})\frac1\bo tr(F_{[ab}F_{cd]}) $$

\subsect Future

Our model provides a tool to better understand certain features of string theory, and how they can be obtained outside of string theory, but we have left several questions unanswered:  This model seems to be nonrenormalizable (as is ordinary supersymmetric Yang-Mills in $D=10$), at least in the usual power-counting sense; in string theory, this problem is resolved by its symmetries.  In particular, we have not determined the exact form of the higher-derivative interactions.  Is this corrected by restrictions from other quantum contributions or supersymmetry?  Specifically, is it possible to reproduce bound-state graviton self-interactions consistent with general relativity?  If so, will we find the proposed energy-momentum anomaly?  And can the usual hexagon anomalies be canceled with our U(N) groups, or is there a way to introduce SO(32)?

\bigskip\noindent{\sectfont \rgbo{0 .5 1}{Acknowledgments
	}}\par\nobreak\medskip

W.S. was supported in part by National Science Foundation Grant No.\ PHY-0098527,  and thanks Radu Roiban for some references and discussions on noncommutative field theory.

\refs

\itemize{1} W. Siegel and B. Zwiebach, {\it Nucl. Phys. B} {\bf 282} (1987) 125;\\
W. Siegel, {\it Nucl. Phys. B} {\bf 284} (1987) 632;\\
N. Berkovits, \xxxlink{hep-th/9607070}, {\it Phys. Lett. B} {\bf 388} (1996) 743.

\itemize{2} J.L. Gervais and A. Neveu, {\it Nucl. Phys. B} {\bf 46} (1972) 381.

\itemize{3} Z. Bern, L. Dixon, D.A. Kosower, \xxxlink{hep-ph/9602280}, {\it Ann. Rev. Nucl. Part. Sci.} {\bf 46} (1996) 109;\\
M.J. Strassler, \xxxlink{hep-ph/9205205}, {\it Nucl. Phys. B} {\bf 385} (1992) 145.

\itemize{4} L. Maldacena, \xxxlink{hep-th/9711200}, {\it Adv. Theor. Math. Phys.} {\bf 2} (1998) 231;\\
S.S. Gubser, I.R. Klebanov and A.M. Polyakov, \xxxlink{hep-th/9802109}, {\it Phys. Lett. B} {\bf 248} (1998) 105;\\
E. Witten, \xxxlink{hep-th/9802150}, {\it Adv. Theor. Math. Phys.} {\bf 2} (1998) 253.

\itemize{5} W. Siegel, \xxxlink{hep-th/9312117}, {\it Phys. Rev. D} {\bf 49} (1994) 4144.

\itemize{6} S. Weinberg and E. Witten, {\it Phys. Lett. B} {\bf 96} (1980) 59.

\itemize{7} W. Siegel, {\it Phys. Lett. B} {\bf 142} (1984) 276.

\itemize{8} H.B. Nielsen and P. Olesen, {\it Phys. Lett. B} {\bf 32} (1970) 203;\\
D.B. Fairlie and H.B. Nielsen, {\it Nucl. Phys. B} {\bf 20} (1970) 637;\\
B. Sakita and M.A. Virasoro, {\it Phys. Rev. Lett.} {\bf 24} (1970) 1146;\\
F. David, {\it Nucl. Phys. B} {\bf 257 [FS14]} (1985) 543;\\
V.A. Kazakov, I.K. Kostov, and A.A. Migdal, {\it Phys. Lett. B} {\bf 157} (1985) 295;\\
J. Ambj\o rn, B. Durhuus, and J. Fr\" ohlich, {\it Nucl. Phys. B} {\bf 257} (1985) 433;\\
M.R. Douglas and S.H. Shenker, {\it Nucl. Phys. B} {\bf 335} (1990) 635;\\
D.J. Gross and A.A. Migdal, {\it Phys. Rev. Lett.} {\bf 64} (1990) 127;\\
E. Br\' ezin and V.A. Kazakov, {\it Phys. Lett. B} {\bf 236} (1990) 144.

\itemize{9} W. Siegel, \xxxlink{hep-th/9601002}, {\it Int. J. Mod. Phys. A} {\bf 13} (1998) 381.

\itemize{10} W. Siegel, {\it Introduction to string field theory} (World Scientific, 1988, Singapore), \xxxlink{hep-th/0107094}, eq. (9.3.2).

\itemize{11} H.S. Snyder, {\it Phys. Rev.} {\bf 71} (1947) 38, {\bf 72} (1947) 68;\\
C.N. Yang, {\it Phys. Rev.} {\bf 72} (1947) 874.

\itemize{12} A. Connes and M.A. Rieffel, {\it Contemp. Math. Oper. Alg, Math. Phys.} {\bf 62} (1987) 237;\\
T. Filk, {\it Phys. Lett. B} {\bf 376} (1996) 53;\\
A. Connes, M.R. Douglas, and A. Schwarz, \xxxlink{hep-th/9711162}, {\it JHEP} {\bf 9802} (1998) 3;\\
N. Seiberg and E. Witten, \xxxlink{hep-th/9908142}, {\it JHEP} {\bf 9909} (1999) 32;\\
M. Li and Y.-S. Wu, {\it Physics in noncommutative world, I: Field theories} (Rinton, 2002, Princeton).

\itemize{13} S. Minwala, M. Van Raamsdonk, and N. Seiberg, \xxxlink{hep-th/9912072}, {\it JHEP} {\bf 0002} (2000) 20;\\
M. Van Raamsdonk and N. Seiberg, \xxxlink{hep-th/0002186}, {\it JHEP} {\bf 0003} (2000) 35.

\itemize{14} H. Liu and J. Michelson, \xxxlink{hep-th/0004013}, {\it Phys. Rev. D} {\bf 62} (2000) 66003,\\
\xxxlink{hep-th/0008205}, {\it Nucl. Phys. B} {\bf 614} (2001) 279;\\
Y. Kiem, S. Lee, and J. Park, \xxxlink{hep-th/0008002}, {\it Nucl. Phys. B} {\bf 594} (2001) 169;\\
Y. Kiem, S.-J. Rey, H.-T. Sato, and J.-T. Yee, \xxxlink{hep-th/0106121}, {\it Phys. Rev. D} {\bf 65} (2002) 26002, \xxxlink{hep-th/0107106}, {\it Eur. Phys. J. C} {\bf 22} (2002) 757;\\
Y. Kiem, S.-S. Kim, S.-J. Rey, and H.-T. Sato, \xxxlink{hep-th/0110066}, {\it Nucl. Phys. B} {\bf 641} (2002) 256;\\
Y. Kiem, S. Lee, S.-J. Rey, and H.-T. Sato, \xxxlink{hep-th/0110215}, {\it Phys. Rev. D} {\bf 65} (2002) 46003.

\itemize{15}  A.H. Chamseddine, {\it Phys. Rev. D} {\bf 24} (1981) 3065;\\
S.J. Gates, Jr. and H. Nishino, {\it Phys. Lett. B} {\bf 157} (1985) 157.

\bye